\documentclass[showpacs,twocolumn,floats,superscriptaddress]{revtex4}
\usepackage{graphicx}
\usepackage{amsmath}
\usepackage{bm}
\usepackage{color}

\newcommand{\gtrsim}{\mbox{\raisebox{2.7pt}{$>$}\hskip -9pt
\raisebox{-2.6pt}{$\sim$}}}

\newcommand{\pll}{\parallel}

\newcommand{\e}{{\rm e}}
\newcommand{\rmd}{{\rm d}}
\newcommand{\rmi}{{\rm i}}

\newcommand{\half}{{\textstyle{\frac{1}{2}}}}

\newcommand{\de}{\delta}

\newcommand{\eps}{\epsilon}

\newcommand{\ignore}[1]{\relax}
\newcommand{\tD}{\tau_{\rm D}}
\newcommand{\tE}{\tau_{\rm E}}
\newcommand{\tEo}{\tau_{\rm E}^{\rm o}}

\newcommand{\NL}{N_{\rm L}}
\newcommand{\NR}{N_{\rm R}}
\newcommand{\WL}{W_{\rm L}}
\newcommand{\WR}{W_{\rm R}}
\newcommand{\gD}{g_{\rm D}}

\begin{document}
\title{Shot noise in semiclassical chaotic cavities}
\author{Robert S. Whitney}
\affiliation{
   D\'epartement de Physique Th\'eorique,
   Universit\'e de Gen\`eve, CH-1211 Gen\`eve 4, Switzerland}
\affiliation{
   Institut Laue-Langevin,
   6 rue Jules Horowitz, BP 156, 38042 Grenoble, France
   }
\author{Ph.~Jacquod}
\affiliation{
   Physics Department, 
   University of Arizona, 1118 E. 4$^{\rm th}$ Street, Tucson, AZ 85721, USA}

\date{April 21, 2006}
\begin{abstract}
We construct a trajectory-based semiclassical theory of shot noise in
clean chaotic cavities. In the universal regime of vanishing Ehrenfest
time $\tE$, we reproduce the random matrix theory result, and show that the
Fano factor is exponentially suppressed as $\tE$ increases. We demonstrate how
our theory preserves the unitarity of the scattering matrix even in the
regime of finite $\tE$. We discuss the range of validity of our
semiclassical approach and point out subtleties relevant 
to the recent semiclassical treatment of shot noise in the universal 
regime by Braun et al. [cond-mat/0511292].
\end{abstract}
\maketitle

{\bf Introduction.}
Quantum transport through chaotic ballistic cavities is often well described 
by Random Matrix Theory (RMT) \cite{carlormt}. Despite its many
successes, or should we say, because of these successes, one might
wonder what is the origin of this RMT universality,
and under what conditions do system specificities
modify the RMT of transport.
System specific contributions to transport originate from
the underlying classical dynamics, which suggests that one employs
semiclassical methods based on classical trajectories \cite{Ric97}.
Indeed,
the semiclassical program toward a microscopic foundation for
the RMT of transport, including explicit bounds for its
regime of applicability, is currently on its way to being
completed successfully \cite{richter,inanc,schanz-fano,wj2004,jw2005-class,haake-weakloc,haake-fano}.

\begin{figure}
\centerline{\hbox{\includegraphics[width=8.4cm]{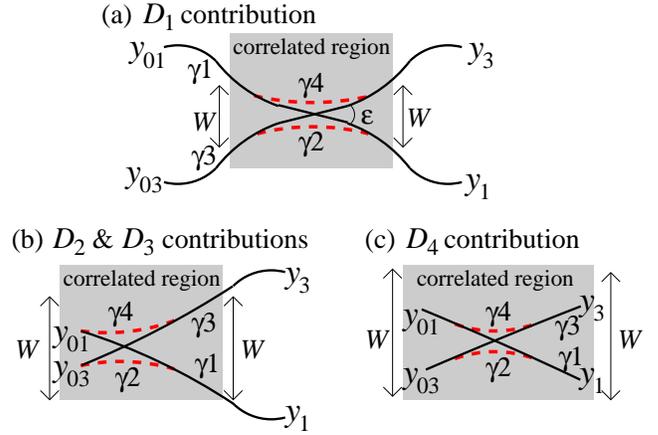}}}
\caption[]{\label{fig1} (Color online) The four dominant contributions to 
${\rm Tr}[{\bf t}^\dagger {\bf t}{\bf t}^\dagger {\bf t}]$.
Paths are paired everywhere except at encounters where
two of them ($\gamma1,\gamma3$)
cross each other (solid lines) while the other two ($\gamma2,\gamma4$)
avoid the crossing (dashed lines).  
(a) Contribution $D_1$ has uncorrelated escape 
on both sides of the encounter. 
(b) Contribution $D_2$ and $D_3$ have
correlated escape only on one side of the encounter.
(c) Contribution  $D_4$ has correlated escape 
on both sides.}
\end{figure}

Here we contribute to this program by deriving the zero-frequency shot noise
power $S$ for quantum chaotic systems. The interest in shot-noise, the 
intrinsically quantum part of the fluctuations of a non-equilibrium electronic 
current, is that it
often contains information on the system that cannot be obtained through
conductance measurements. For instance, shot-noise experiments have 
determined the charge and statistics of the charge carriers in
superconducting heterostructures or in the fractional quantum hall 
effect \cite{BlanterPR}. In this paper, we consider an open ballistic
quantum dot \cite{marcus} carrying a large number of conducting channels
and accordingly neglect electron-electron interactions.
We reproduce the RMT result, and 
show how shot noise deviates from RMT predictions in the semiclassical limit.
We calculate the Fano factor $F=S/S_{\rm p}$,
given by the ratio of $S$ to the Poissonian noise 
$S_{\rm p}=2 e \langle I \rangle$ that would be generated by a current 
flow of uncorrelated electrons. According to the scattering theory of
transport one has 
$F={\rm Tr}[{\bf t}^\dagger {\bf t}(1-{\bf t}^\dagger {\bf t})]/
{\rm Tr}[{\bf t}^\dagger {\bf t}]$ \cite{BlanterPR}.
If one makes the RMT assumption that
the transmission matrix ${\bf t}$ is the $\NL \times \NR$ off-diagonal 
block of 
a $(\NL + \NR)\times (\NL + \NR)$ random unitary scattering
matrix, one gets
$F=\NL\NR/(\NL+\NR)^2$ \cite{carlormt,BlanterPR},
in term of the number of quantum channels $\NL$ and $\NR$ carried 
by the contacts to the left and right leads. 
Ref.~\cite{schanz-fano} carried out 
the first semiclassical calculation of $F$ 
for the specific case of quantum graphs. The difficulty is to calculate 
${\rm Tr}[{\bf t}^\dagger {\bf t} {\bf t}^\dagger {\bf t}] = \sum_{i,j,q}
|t_{j,i}|^2 |t_{q,i}|^2 + \sum_{i,j,p} |t_{j,i}|^2 |t_{j,p}|^2 
+ \sum_{i \ne p;j \ne q} t^*_{j,i} t_{j,p} t^*_{q,p} t_{q,i}$.
Ref.~\cite{schanz-fano} employed a diagonal approximation to calculate the
first two terms and identified the dominant four-trajectory contributions to
the third one. Quantum graphs fundamentally 
differ from continuum models which we treat here. 
In our semiclassical derivation
we find the dominant contributions to $F$
from the path pairings shown in Fig.~\ref{fig1}.
These pairings are similar to those considered
in Ref.~\cite{schanz-fano} for quantum graphs.
However, unlike quantum graphs, chaotic systems have continuous families 
of scattering trajectories with similar actions, which
means in particular that we cannot make a diagonal approximation 
to evaluate the contributions $D_2$ and $D_3$ shown in Fig.~\ref{fig1}. 
This important point was not addressed in Ref.~\cite{haake-fano}.

Exploring the range of validity of RMT for
chaotic systems, we find 
$F$ to be exponentially reduced \cite{agam},
\begin{eqnarray}
F = \NL\NR (\NL+\NR)^{-2} \exp [-\tEo/\tD],
\label{eq:Fano-result}
\end{eqnarray} 
for systems with left (right) lead width, $\WL$ ($\WR$), such that 
the width of leads
$W_{\rm L,R} \; \gtrsim \; \hbar_{\rm eff}^{1/2} L$. These systems
witness the emergence of the new Ehrenfest time scale 
$\tEo = \lambda^{-1} \ln [\hbar_{\rm eff}^{-1}(\tau_{\rm f}/\tD)^2]$,
which generically induces significant deviations 
from the RMT of transport \cite{Scho05}.
Here, $\hbar_{\rm eff}=\hbar/(p_{\rm F}L)$,
$L$ is the linear system size, $p_{\rm F}$ the Fermi momentum
of the particle with mass $m$,
$\tau_{\rm f}$ the time of flight, $\tD$ the dwell time through the
system, 
and $\lambda$ the Lyapunov exponent of the chaotic classical
dynamics. 

Our semiclassical calculation correctly captures both the
universal regime with $\tEo/\tD \ll 1$ and the deep semiclassical
regime where $\tEo$ becomes comparable to or exceeds $\tD$. We reproduce
Eq.~(\ref{eq:Fano-result}) and explicitly show 
that the exponential suppression of $F$ is due to 
paths shorter than $\tEo$ which become noiseless
\cite{wj2004,jw2005-class,silvestrov}.
We demonstrate the unitarity of the
theory by calculating both 
$F={\rm Tr}[{\bf t}^\dagger {\bf t}(1-{\bf t}^\dagger {\bf t})]/{\rm Tr}
[{\bf t}^\dagger {\bf t}]$ 
and $F={\rm Tr}[{\bf t}^\dagger {\bf t} {\bf r}^\dagger {\bf r}]/{\rm Tr}
[{\bf t}^\dagger {\bf t}]$. We finally comment on the current limitations
of the trajectory-based semiclassical approach.

We consider a two-dimensional chaotic quantum dot ideally connected 
to two external leads. We require that the size 
of the openings to the leads is much smaller than the perimeter of the system 
but is still semiclassically large, $1 \ll \NL,\NR \ll L/\lambda_F$.
This ensures that the chaotic dynamics inside the dot has enough time to 
develop. 
The system's transport properties are given by its
scattering matrix ${\cal S}$, with
an $\NL \times \NR$ transmission block ${\bf t}$,
and an  $\NL \times \NL$ reflection block ${\bf r}$. 
To calculate the Fano factor, one needs to calculate
the conductance $g = {\rm Tr} [{\bf t}^\dagger {\bf t}]$, as well as
${\rm Tr}[{\bf t}^\dagger{\bf t}{\bf t}^\dagger{\bf t}]$.


Semiclassically, the transmission matrix reads \cite{Bar93},
\begin{eqnarray}
t_{ji} 
&=&
-(2\pi \rmi \hbar)^{-1/2}
\!\int_{\rm L} \! \! \rmd y_0 \int_{\rm R} \! \rmd y 
\sum_\gamma (\rmd p_y/\rmd y_0)^{1/2}_\gamma
\nonumber \\
& &\qquad \times \langle j|y\rangle  
\langle y_0| i \rangle 
\exp[\rmi S_\gamma /\hbar + \rmi \pi \mu_\gamma /2]
\, ,\qquad
\end{eqnarray} 
where $|i\rangle$ is the transverse wavefunction
of the $i$th lead mode. This expression 
sums over all paths $\gamma$ (with classical action $S_{\gamma}$ 
and Maslov index $\mu_\gamma$) starting at $y_0$ on a cross-section of
the injection (L) lead and ending at $y$ on the
exit (R) lead. We approximate 
$\sum_n\langle y'|n\rangle\langle n|y\rangle \approx 
\delta (y'-y)$ \cite{footnote:delta_hbar},
to write
${\rm Tr}[{\bf t}^\dagger {\bf t}{\bf t}^\dagger {\bf t}]$
as a sum over four paths, $\gamma1$ from $y_{01}$ to $y_1$,
$\gamma2$ from $y_{03}$ to $y_1$,
$\gamma3$ from $y_{03}$ to $y_3$ and
$\gamma4$ from $y_{01}$ to $y_3$, 
\begin{eqnarray}\label{trt4}
{\rm Tr} [{\bf t}^\dagger {\bf t}{\bf t}^\dagger {\bf t}]
&=& 
{1\over (2\pi \hbar)^2}
\!\int_{\rm L} \! \! \rmd y_{01} \rmd y_{03} \int_{\rm R} \! \rmd y_1 \rmd y_3  \nonumber \\
& \times &
\sum_{\gamma1,\cdots \gamma4} 
A_{\gamma4}A_{\gamma3} A_{\gamma2}A_{\gamma1}
\exp [\rmi\delta S/\hbar] \,.
\end{eqnarray}
Here, $A_\gamma = [\rmd p_y/\rmd y_0]_\gamma^{1/2}$
and $\delta S = S_{\gamma1}-S_{\gamma2}+S_{\gamma3}-S_{\gamma4}$
(we absorbed all Maslov indices into the actions
$S_{\gamma i}$).
We are interested in quantities 
averaged over variations in the energy or the system shape. 
For most contributions, $\delta S/\hbar$
oscillates wildly with these variations. 
The dominant contributions that survive averaging are those for which
the fluctuations of $\delta S/\hbar$ are minimal. They are shown in 
Fig.~\ref{fig1}. 
Their paths are in pairs almost everywhere except in the
vicinity of encounters. Going through an encounter,
two of the four paths cross each other, while the other two
avoid the crossing. They remain in pairs, though the pairing switches,
e.g. from $(\gamma1;\gamma4)$ and $(\gamma2;\gamma3)$ to 
$(\gamma1;\gamma2)$ and $(\gamma3;\gamma4)$ in Fig.~\ref{fig1}a.
Paths are always close enough to their partner
that their stability is the same.
Thus, for all pairings in
Fig.~\ref{fig1},
\begin{equation}\label{paths-pairing}
\sum_{\gamma1,...\gamma4} A_{\gamma4}A_{\gamma3} A_{\gamma2}A_{\gamma1}
\rightarrow \sum_{\gamma1,\gamma3} A_{\gamma3}^2 A_{\gamma1}^2.
\end{equation}
We define $P({\bf Y},{\bf Y}_0;t)\de y\de \theta \de t$
as the product of the momentum along the injection lead,
$p_{\rm F}\cos \theta_0$, and the classical probability to go
from an initial position and angle
${\bf Y}_0=(y_0,\theta_0)$ to within $(\de y,\de \theta)$ of
${\bf Y}$ in a time within $\de t$ of $t$. 
Then the sum over
all paths $\gamma$ from  $y_0$ to $y$ is
\begin{eqnarray}
\sum_\gamma \!
A^2_\gamma \;
\! [\cdots]_\gamma 
\!\! &=& \!\!
\int_0^\infty \! \rmd t \! \int \! \rmd \theta_0  \! \int \! \rmd \theta 
\; P({\bf Y},{\bf Y}_0;t) 
\; [\cdots]_{{\bf Y}_0}.
\quad
\label{eq:gamma-sum-to-Pintegral}
\end{eqnarray}
For an individual system, $P$ has $\de$-functions for
all classical trajectories.  However averaging over 
an ensemble of systems or over energy 
gives a smooth function
\begin{eqnarray}
\langle P({\bf Y},{\bf Y}_0;t) \rangle = \frac{
p_{\rm F} \cos \theta_0 \cos \theta }
{2 (W_{\rm L}+W_{\rm R})\tau_{\rm D}} \; 
\exp[-t/\tau_{\rm D}] \, .
\label{eq:average-P}
\end{eqnarray}
Using Eqs.~(\ref{eq:gamma-sum-to-Pintegral}) and (\ref{eq:average-P}) 
to calculate the conductance within the diagonal approximation 
directly leads to the Drude conductance
$\langle {\rm Tr} [{\bf t}^\dagger {\bf t}] \rangle 
\simeq g_{\rm D} = \NL \NR/(\NL+\NR)$. 
This level of approximation for $\langle {\rm Tr} [{\bf t}^\dagger {\bf t}]
\rangle $
is sufficient to obtain $F$ to leading order
in $N_{\rm L,R}^{-1}$.
We now use Eqs.~(\ref{trt4}), (\ref{paths-pairing})
and (\ref{eq:gamma-sum-to-Pintegral}) to
analyze the contributions in
Fig.~\ref{fig1}.

There are two things that can happen to two pairs of paths
as they leave an encounter. The first is {\it uncorrelated escape}. 
The pairs of paths escape when
the perpendicular distance between them is larger than $W_{\rm L,R}$,
which requires a minimal time
$T_W(\eps)/2 = \lambda^{-1} \ln [\eps^{-1} W/L]$ 
between encounter and escape. The two pairs of paths
then escape in an uncorrelated manner, typically at completely different
times, with completely different momenta (and possibly through 
different leads).
The second is {\it correlated escape}.
Pairs of paths escape when the distance between them is less 
than $W_{\rm L,R}$, 
then the two pairs of paths escape together, at the same
time through the same lead. 

{\bf Contributions to the Fano factor.}
Taking into account the two escape scenarios just described, 
we write $\langle {\rm Tr}[{\bf t}^\dagger {\bf t}{\bf t}^\dagger {\bf t}]
\rangle = D_1+D_2+D_3+D_4$. Each of these four contributions, sketched in 
Fig.~\ref{fig1}, can be written as
\begin{eqnarray}\label{contribution}
D_i 
=
{1\over (2\pi \hbar)^2}
\!\int_{\rm L} \! \! \rmd {\bf Y}_{01} \; \rmd {\bf Y}_{03} 
\!\int_{\rm R} \! \! 
\rmd {\bf Y}_{1} \; \rmd {\bf Y}_{3} \!\int \! \! \rmd t_1 \; \rmd t_3 
\nonumber \\
\times \;
\langle P({\bf Y}_1,{\bf Y}_{01};t_1) \;  P({\bf Y}_3,{\bf Y}_{03};t_3) 
\rangle \; \exp [\rmi\delta S_{D_i}/\hbar] \,,
\end{eqnarray}
where subscripts ${1,3}$ make the connection to Fig.~\ref{fig1}. 
When evaluating Eq.~(\ref{contribution}) the joint exit probability for two
crossing paths has to be computed.

To evaluate $D_1$, we use the method developed by Richter and Sieber 
\cite{richter}, 
while 
taking into account
that paths in the same region of phase-space 
(shaded areas in Fig.~\ref{fig1})
have highly correlated escape probabilities \cite{Bro05}.
Here the action difference is 
$\delta S_{D_1}= E_{\rm F}\eps^2/\lambda$ \cite{richter}, 
where $\eps$ is the crossing angle shown in Fig.~\ref{fig1}a.
We write
\begin{equation}
P({\bf Y}_i,{\bf Y}_{0i};t_i)= \!
\int \rmd {\bf R}_i 
\tilde{P}({\bf Y}_i,{\bf R}_i;t_i-t_i')P({\bf R}_i,{\bf Y}_{0i};t_i') \, ,
\nonumber
\end{equation} 
where $\tilde{P}$ is the probability for the classical path to exist 
(not multiplied by the injection momentum), and
${\bf R}_i$ is a point in the system's phase-space 
$({\bf r}_i,\phi_i)$ visited at time $t_i'$, 
with $\phi_i$ giving the direction of the momentum.
We choose ${\bf R}_1$ and ${\bf R}_3$ as the points at which the paths 
cross, so ${\bf R}_3 = ({\bf r}_1,\phi_1 \pm \eps)$ and
$\rmd {\bf R}_3 = v_{\rm F}^2 \sin \eps \rmd t_1' \rmd t_3' \rmd \eps$. 
Thus
\begin{eqnarray}
D_1 
&=& 
2(2\pi \hbar)^{-2}  \int_{\rm L} \rmd {\bf Y}_{01} \rmd {\bf Y}_{03}
\nonumber \\
& \times & \int_0^\pi \rmd \eps \;
{\rm Re}\big[\e^{\rmi \de S_{D_1}/\hbar}\big] 
\big\langle I( {\bf Y}_{01}, {\bf Y}_{03};\eps) \big\rangle .
\label{eq:D_1-as-integral}
\end{eqnarray}
$I( {\bf Y}_{01}, {\bf Y}_{03};\eps)$ 
is related to the probability that $\gamma3$ crosses $\gamma1$ 
at angle $\pm\eps$. Its average is independent of
${\bf Y}_{01,03}$, so 
$\langle I( {\bf Y}_{01}, {\bf Y}_{03};\eps) \rangle = 
\langle I(\eps)\rangle$.
For $D_1$, injections/escapes are more than $T_W(\eps)/2$ from the crossing, so
\begin{eqnarray}
\langle I(\eps)\rangle
\! &=& \! 
2v_{\rm F}^2 \; \sin \eps
\int_{\rm R} \rmd {\bf Y}_{1} \rmd {\bf Y}_{3} \int \rmd {\bf R}_1 \nonumber \\
& \times & \int_T^\infty \rmd t_1  
\int_{T/2}^{t_1-T/2} \! \rmd t_1' 
\int_T^\infty \rmd t_3 
\int_{T/2}^{t_3-T/2} \! \rmd t'_3
\nonumber \\
& \times & \,
\big\langle \tilde{P}({\bf Y}_1,{\bf R}_1;t_1-t_1') 
P({\bf R}_1,{\bf Y}_{01};t_1') 
\nonumber \\
& \times & \, 
\tilde{P}({\bf Y}_3,{\bf R}_3;t_3-t_3') 
P({\bf R}_3,{\bf Y}_{03};t_3') \big\rangle  \, ,
\quad \quad 
\end{eqnarray}
where $T$ is shorthand for $T_W(\eps)$. We next note that
within $T_W(\eps)/2$ of the crossing, paths $\gamma1$ and $\gamma3$
are so close
to each other that their joint escape probability is 
the same as for a single path
(this was absent from Ref.~\cite{richter} and was first noted 
in Ref.~\cite{Bro05}). 
Elsewhere $\gamma1,\gamma3$ escape
independently through either lead at anytime, hence 
\begin{eqnarray}
\big\langle I(\eps) \big\rangle
=
{p_{\rm F}^4  \tau_{\rm D} \over \pi \hbar m }
{ N_{\rm R}^2 
\cos \theta_{01}\cos\theta_{03} \sin \epsilon 
\over  (N_{\rm L}+N_{\rm R})^3} \;
e^{-T_W(\eps)/\tau_{\rm D}} \, , \ \ 
\end{eqnarray}
where we used $\NR=(\pi \hbar)^{-1}p_{\rm F}\WR$, 
and assumed that the probability that $\gamma3$ is at 
${\bf R}_3$ at time $t_3'$ 
in a system of area $A$ is 
$(2\pi A)^{-1} =  m [2\pi\hbar \tau_{\rm D}(\NL+\NR)]^{-1}$.
Then the ${\bf Y}_{01,03}$-integral in Eq.~(\ref{eq:D_1-as-integral})
gives 
$(2\WL)^{2}$, while the $\eps$-integral 
is dominated by $\eps \ll 1$ and yields a factor of
$-\pi \hbar(2E_{\rm F} \tau_{\rm D})^{-1} \e^{-\tEo/\tD}
\big\{1+ {\cal O}[(\lambda \tD)^{-1}]\big\}$ 
\cite{inanc}. Thus
\begin{eqnarray}\label{D1}
D_1 &=& -\NL^2 \NR^2(\NL+\NR)^{-3} \exp[-\tEo/\tD] \, .
\end{eqnarray}

\begin{figure}
\centerline{\hbox{\includegraphics[width=8.5cm]{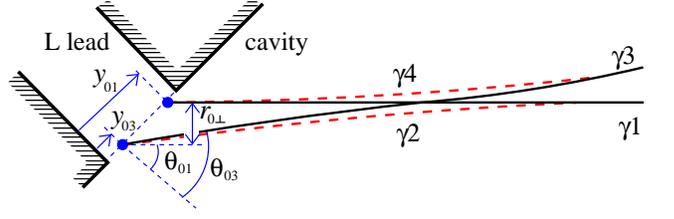}}}
\caption[]{\label{fig-D2} (Color online)
Paths for the $D_2$ and $D_4$ contributions when they are in the
correlated region (close to L lead).
Paths $\gamma 1$ and $\gamma3$ (solid black lines)
start on the cross-section of the L lead at positions
$y_{01}$ and $y_{03}$ with transverse momenta 
$p_{\rm F} \sin \theta_{01}$ and
$p_{\rm F} \sin \theta_{03}$, respectively.
In the basis parallel/perpendicular to $\gamma1$, 
the initial position and momentum of path $\gamma3$ are
$r_{0\perp} = (y_{01}-y_{03})\cos \theta_{01} $, 
$r_{0\pll} = (y_{01}-y_{03})\sin \theta_{01}$  
and $p_{0\perp} \simeq p_{\rm F} (\theta_{01}-\theta_{03})$.
Contribution $D_3$ has exactly the same structure
close to the R lead.
}
\end{figure}

The contribution $D_2$ is shown in Fig.~\ref{fig1}b, with
Fig.~\ref{fig-D2} showing the paths in the correlated region in more
detail.
Noting that $\gamma2$ decays exponentially towards $\gamma1$, we find
the action difference between the two paths to be
\begin{eqnarray}
S_2-S_1 &=& p_{\rm F}(y_{01}-y_{03})\sin \theta_{01} 
\nonumber\\
& &\qquad + \half m\lambda (y_{01}-y_{03})^2\cos^2 \theta_{01} \, . \quad
\label{eq:correlated-action-diff}
\end{eqnarray}
The equation for 
$S_4-S_3$ has the opposite sign for $(y_{01}-y_{03})$
and $\theta_{01}$ replaced by $\theta_{03}$.
In terms of $(r_{0\perp},p_{0\perp})$
\begin{eqnarray}
\de S_{D_2} 
&=& -(p_{0\perp}+m\lambda r_{0\perp}) \, r_{0\perp}\, ,
\end{eqnarray}
where we have dropped cubic terms (they only give $\hbar$-corrections 
to the stationary-phase integral). We next perform the
average in Eq.~(\ref{contribution}).
We define $T'_W(r_{0\perp}, p_{0\perp})$ 
as the time for which $\gamma1$ and $\gamma3$ 
are less than $W$ apart, and insist
that the paths are more than
$W$ apart before they escape to the right. 
Hence we must evaluate
\begin{eqnarray}
& & \hskip -4mm \int_{\rm R} \! \rmd {\bf Y}_1  \rmd {\bf Y}_3
\int_{T'_W}^\infty \rmd t_1 \rmd t_3
\langle P({\bf Y}_1,{\bf Y}_{01};t_1) P({\bf Y}_3,{\bf Y}_{03};t_3) \rangle 
\nonumber \\
& &
= {N_{\rm R}^2 p_{\rm F}^2 \cos \theta_{01} \cos \theta_{03} 
\over (N_{\rm L} +N_{\rm R})^2} \; 
\exp[-T'_W(r_{0\perp}, p_{0\perp})/\tau_{\rm D}] \  
\, .\qquad \ 
\end{eqnarray}
Inserting  this into Eq.~(\ref{contribution}),
we change integration variables
using 
$p_{\rm F} \cos \theta _{03}\rmd {\bf Y}_{03} 
=\rmd r_{0\perp} \rmd p_{0\perp}$
\cite{Bar93}, and then define 
$\tilde{p}_0\equiv p_{0\perp}+ m\lambda r_{0\perp}$.
In the regime of interest 
$T'_W(r_{0\perp}, p_{0\perp}) 
\simeq \lambda^{-1} \ln[(m\lambda W)^{-1}\tilde{p}_0]$.
Evaluating the integral over $r_{0\perp}$ leaves 
a $\tilde{p}_0$-integral which we cast as Euler $\Gamma$-functions.
To lowest order in $(\lambda\tD)^{-1}$ we find,
\begin{eqnarray}\label{D2}
D_2
&=& \NL \NR^2 (\NL+\NR)^{-2} \exp[-\tEo/\tD] \, .
\label{eq:D2}
\end{eqnarray}
Substituting
$\NL\leftrightarrow \NR$ in the derivation of Eq.~(\ref{eq:D2}) gives, 
\begin{eqnarray}\label{D3}
D_3
&=& \NL^2\NR (\NL+\NR)^{-2} \exp[-\tEo/\tD]  \, . \qquad
\end{eqnarray}

The contribution $D_4$ is shown in Fig.~\ref{fig1}c,
with Fig.~\ref{fig-D2} showing the paths in detail at the L lead.
This contribution can be evaluated in a way similar to $D_2$,
the difference being that the paths escape before time 
$T'_W(r_{0\perp},p_{0\perp})$, i.e. before becoming a distance
$W$ apart.  The paths are always correlated, so the escape probability
for the two paths equals that for one. Moreover,
both paths will automatically escape through the same lead,
hence
\begin{eqnarray}
& &\int_{\rm R} \! \rmd {\bf Y}_1  \rmd {\bf Y}_3 
\int_0^{T'_W} \!\!\! \rmd t_1\rmd t_3 \,
\big\langle P({\bf Y}_1,{\bf Y}_{01};t_1) P({\bf Y}_3,{\bf Y}_{03};t_3)
\big\rangle
\nonumber \\
& &\quad = {N_{\rm R} p_{\rm F}^2 \cos \theta_{01} \cos \theta_{03} 
\over N_{\rm L} +N_{\rm R}}
\big( 1-\e^{-T'_W(r_{0\perp}, p_{0\perp})/\tD} \big) \, . 
\end{eqnarray}
Performing the same analysis as for $D_2$ we find that
\begin{eqnarray}\label{D4}
D_4
&=& \NL\NR(\NL+\NR)^{-1} (1-\exp[-\tEo/\tD]) \, . \qquad 
\end{eqnarray}
The Fano factor is given by $F= 1-\gD^{-1}(D_1+D_2+D_3+D_4)$.
Our results of Eqs.~(\ref{D2}), (\ref{D3}) and (\ref{D4}) show that
$D_2+D_3+D_4=g_{\rm D}$. One hence gets $F= -D_1/g_{\rm D}$.
From Eq.~(\ref{D1}), one finally obtains our 
main result, Eq.~(\ref{eq:Fano-result}).
The splitting of phase-space discussed in 
Refs.~\cite{silvestrov,wj2004,jw2005-class} for $\tEo \gtrsim \tD$
naturally emerges here. 
For paths shorter than $\tEo$, only $D_4$ is non-zero.
This cancels these path's ${\rm Tr} [{\bf t}^\dagger {\bf t}]$-contribution,
making them noiseless.

{\bf Preservation of Unitarity.}
The unitarity of the scattering matrix 
ensures that ${\bf t}^\dagger{\bf t}+{\bf r}^\dagger{\bf r}=1$
and hence the Fano factor can be written as
$F=\gD^{-1}\langle{\rm Tr}[{\bf t}^\dagger{\bf t}{\bf r}^\dagger{\bf r}]
\rangle$. We calculate this expression to explicitly
show that our method preserves unitarity.
We first note that there is no contribution $D_3$ nor $D_4$
to ${\rm Tr}[{\bf t}^\dagger{\bf t}{\bf r}^\dagger{\bf r}]$.
We are left with the calculation of two contributions,
$D_1'$ and $D_2'$, obtained from $D_1$ and $D_2$ shown in 
Fig.~\ref{fig1}a,b
with $y_{01},y_{03}$ and $y_3$ on the left lead  and $y_1$ on the right lead.
The calculation proceeds as for $D_1$ and $D_2$, with one factor of 
$\NR/(\NL+\NR)$ replaced by $\NL/(\NL+\NR)$ in both contributions.
The sum of these two contributions is 
$D_1'+D_2'= \e^{-\tEo/\tD} \NL^2 \NR^2(\NL+\NR)^{-3}$,
the Fano factor is then $F=(D_1'+D_2')/g_{\rm D}$, which reproduces
Eq.~(\ref{eq:Fano-result}).

{\bf Off-diagonal nature of all contributions.}
In our analysis we allow for the fact that
open chaotic systems have continuous families of paths 
with highly correlated actions coupling to multiple lead modes.
For example paths $\gamma1$ and $\gamma2$ in Fig.~\ref{fig-D2}
have an action difference given in 
Eq.~(\ref{eq:correlated-action-diff}), which does not fluctuate
under energy or sample averaging.
The stationary phase integral for $D_{2,3,4}$ over such paths is 
dominated by paths $\gamma1$ and $\gamma3$
with $p_{0\perp}\simeq -m\lambda r_{0\perp}$. Since
$r_{0\perp}$ is integrated over the width $W$ of the lead,
$p_{0\perp}$ varies over a range of order $ m\lambda W$, 
these contributions are clearly
{\it not} diagonal in the lead mode basis.
Upon completion of this manuscript, we became aware of Ref.~\cite{haake-fano}, 
which presents a semiclassical calculation of $F$ for $\tEo=0$. 
While their method is superficially similar to ours,
they make a diagonal assumption to get 
the contributions that we call $D_{2,3,4}$.  
Our analysis shows that this is unjustifiable. Such
an assumption would moreover violate unitarity for finite $\tEo$.

{\bf Regime of applicability of these semiclassics.} 
We appear to be the first to report that
all trajectory-based 
semiclassical methods used so far in the theory of transport
(including in the present article)
are only applicable in the regime $W \geq \hbar_{\rm eff}^{1/2}L$.
Dominant off-diagonal contributions
such as those discussed above have encounters of a typical
size $\sim \hbar_{\rm eff}^{1/2} L$. When 
$W < \hbar_{\rm eff}^{1/2}L$,
the two non-crossing paths (i.e. $\gamma2$ and $\gamma4$ in Fig.~\ref{fig1})
at an encounter
are a distance apart greater than $W$. 
The probability that one of the four paths
escapes while the other three paths remain in the
system is of order $\tilde{\tau}_{\rm E}/\tau_{\rm D}$, where 
$\tilde{\tau}_{\rm E} \sim \lambda^{-1} \ln [\hbar_{\rm eff} (L/W)^2] $
is the time over which this path is a distance of order $W$ from any of 
the other paths. 
The current methods fail once this is taken into account,
suggesting that diffraction effects
may become important.
We believe that the regime $\hbar_{\rm eff}<(W/L) \le \hbar_{\rm eff}^{1/2}$
is well described by RMT, and thus
suspect this diffraction may be 
the microscopic source of RMT universality in this regime.
Clearly this regime merits further study.


This work has been financially 
supported by the Swiss National Science Foundation.

\bibliographystyle{apsrev}

\begin{thebibliography}{99}
\bibitem{footnote:delta_hbar}
For an ideal lead with $N$ lead modes of the form 
$\langle y|n\rangle= (2/W)^{1/2} \sin (\pi yn/W)$,
one finds that 
$\sum_n\langle y'|n\rangle\langle n|y\rangle
=(2W)^{-1}[ \sin [(z'-z)(N+1/2)] /\sin [(z'-z)/2]
- \sin [(z'+z)(N+1/2)] /\sin [(z'+z)/2]]$
 where $z=\pi y/W$.
This function is strongly peaked at $y'=y$
with peak width $\sim\lambda_{\rm F}$ and height $\sim\lambda_{\rm F}^{-1}$,
in the semiclassical limit we can calculate all quantities
to lowest order in $\hbar$ by approximating this function with a Dirac 
$\de$-function.
\end{thebibliography}

\end{document}